# Mechanical responses of two-dimensional MoTe$_2$; pristine 2H, 1T and 1T' and 1T'/2H heterostructure

Bohayra Mortazavi[*,1], Golibjon R Berdiyorov[2], Meysam Makaremi[3], Timon Rabczuk[#,4]

[1]*Institute of Structural Mechanics, Bauhaus-Universität Weimar, Marienstr. 15, D-99423 Weimar, Germany.*

[2]*Qatar Environment and Energy Research Institute, Hamad Bin Khalifa University, Qatar Foundation, Doha, Qatar.*

[3]*Department of Materials Science and Engineering, University of Toronto, 184 College Street, Suite 140, Toronto, ON M5S 3E4, Canada.*

[4]*College of Civil Engineering, Department of Geotechnical Engineering, Tongji University, Shanghai, China.*

**Abstract**

Transition metal dichalcogenides (TMD) are currently among the most interesting two-dimensional (2D) materials due to their outstanding properties. MoTe$_2$ involves attractive polymorphic TMD crystals which can exist in three different 2D atomic lattices of 2H, 1T and 1T', with diverse properties, like semiconducting and metallic electronic characters. Using the polymorphic heteroepitaxy, most recently coplanar semiconductor/metal (2H/1T') few-layer MoTe$_2$ heterostructures were experimentally synthesized, highly promising to build circuit components for next generation nanoelectronics. Motivated by the recent experimental advances, we conducted first-principles calculations to explore the mechanical properties of single-layer MoTe$_2$ structures. We first studied the mechanical responses of pristine and single-layer 2H-, 1T- and 1T'-MoTe$_2$. In these cases we particularly analyzed the possibility of engineering of the electronic properties of these attractive 2D structures using the biaxial or uniaxial tensile loadings. Finally, the mechanical-failure responses of 1T'/2H-MoTe$_2$ heterostructure were explored, which confirms the remarkable strength of this novel 2D system.

*Corresponding author (Bohayra Mortazavi): bohayra.mortazavi@gmail.com
Tel: +49 157 8037 8770; Fax: +49 364 358 4511; [#]Timon.rabczuk@uni-weimar.de

# 1. Introduction

The great success of graphene [1,2], emerged the two-dimensional (2D) materials as a new class of materials suitable for numerous and diverse applications ranging from nanoelectronics to aerospace structures. For the applications in nanoelectronics, presenting semiconducting electronic character with moderate and tuneable band-gap is highly desirable. Nevertheless, graphene in its pristine form presents zero-band-gap semiconducting property and such that opening a band-gap in graphene requires complex physical or chemical modifications such as chemical doping or defect engineering [3–5]. As an alternative, synthesize of other 2D materials with inherent semiconducting character such as transition metal dichalcogenides (TMD) [6–9] has been considered as a more reliable approach to exploit in post-silicon electronics.

TMD include a stoichiometric formula of $MX_2$ in which M is a transition metal element; such as Mo, W, Sc, Cr or V; and X is a chalcogen element; like S, Se or Te. A single-layer TMD contains a layer of M atoms sandwiched between two layers of X atoms. These attractive 2D nanostructures can be synthesized using a top-down exfoliation approach from the bulk TMD structures [10], or they can be directly grown by a bottom-up technique such as molecular beam epitaxy (MBE) [11] or chemical vapour deposition (CVD) [12]. The physical properties of TMD nanosheets may show remarkable thickness dependency. In this regards, although the bulk $MoS_2$ structure shows an indirect band-gap semiconducting nature, the single-layer $MoS_2$ interestingly presents a direct band-gap electronic character which is highly desirable for the application in nanoelectronics [13–16]. TMD nanomembranes also exhibit reliable stability and mechanical characteristics with stretchabilities comparable to that of the graphene [17–19]. Notably, one can further tune the direct/indirect electronic band-gap of mono/multi-layer TMD through applying the mechanical loading [20–23]. An exciting fact about the TMD structures is their polymorphism nature. In this regard, TMD nanostructures may exist in three different structural phases including; 2H, 1T and 1T' [24,25]. 2H and 1T phases contain the hexagonal and trigonal structure, respectively, whereas 1T' is a distorted form of 1T. The polymorphism nature of TMD may play critical roles on their physical properties and practical aspects. The $MoX_2$ and $WX_2$ TMD nanosheets exhibit contrasting semiconducting and metallic electronic characters for the 2H and 1T atomic configurations, respectively [26,27]. Among the various TMD, $MoTe_2$ nanosheets have recently attracted remarkable attention because of their very promising electronic, photonic and optical properties [28–30]. Contrasting electronic characteristic of pristine TMD phases, consequently proposes the fabrication of heterostructures made from different phases in seamlessly stitched forms in order to further tune the electronic and other physical properties. In this context, several single-layer TMD heterostructures have been experimentally realized so far, such as 2H/1T $MoS_2$ [31]. Most recently, 1T'/2H-$MoTe_2$ [28], a novel TMD coplanar heterostructure with atomically sharp and seamless interfaces was realized experimentally, which shows very promising performances as a field-effect transistor [28]. In a latest experimental study, the structural phase transition from 2H to 1T' in monolayer $MoTe_2$ was achieved by electrostatic doping [29]. Empante and co-workers [32] successfully employed the CVD method to synthesize 2H, 1T and 1T' pristine phases of $MoTe_2$.

For the design of novel nanodevices using these 2D materials, comprehensive understanding of electronic, mechanical, optical and thermal properties of 2D components play crucial roles [33–44]. For the materials at nanoscale such as the 2D materials, experimental techniques for the evaluation of properties are complicated, time consuming and expensive as well. In these cases, first-principles theoretical approaches can be considered as fast viable alternatives to explore various material properties in low-cost and trustable level of precision. Kan *et al.* [45] explored the phase stability and Raman vibration of the 2H, 1T and 1T' $MoTe_2$ monolayers using the first-principles simulations. Bera *et al.* [46] recently studied the pressure-dependent semiconductor to semimetal in 2H-$MoTe_2$ using the first-principles calculations. The mechanical properties of single-layer 2H-MoTe$_2$ were also investigated using the first-principles methods by Kumar and Ahluwalia [47] and Li et al. [19]. To the best of our knowledge, the mechanical properties of 1T and 1T' MoTe$_2$ monolayers and the possibility of band-gap opening in these systems by the mechanical loading have not been studied. Moreover, only very recently the 1T'/2H coplanar heterostructures of MoTe$_2$ were realized experimentally [28] and such that studying the mechanical properties of these heterostructures is highly attractive. Motivated by the recent experimental advances in the fabrication of MoTe$_2$ pristine phases and heterostructures, the objective of present investigation is to explore the mechanical properties as well as the strain engineering of electronic properties of these novel 2D systems by conducting extensive first-principles density functional theory (DFT) simulations.

## 2. Methods

The DFT calculations were performed using the Vienna ab-initio simulation package (VASP) [48–50]. The plane wave basis set with an energy cut-off of 500 eV and the generalized gradient approximation (GGA) exchange-correlation functional proposed by Perdew-Burke-Ernzerhof [51] were employed.

In this work, we analyzed the mechanical responses of single-layer MoTe$_2$ structures by performing uniaxially tensile loading simulations. We applied periodic boundary conditions along all three Cartesian directions and such that the obtained results represent the properties of large-area single-layer films and not the nanoribbons. Since the dynamical effects such as the temperature are not taken into consideration and periodic boundary conditions were also applied along the planar directions, only a unit-cell modelling is accurate enough for the evaluation of mechanical properties [52], we used a unit-cell to model the mechanical properties of pristine MoTe$_2$ phases (3 atoms for the 2H and 1T phases and 6 atoms for the 1T' structure). We considered a vacuum layer of 20 Å to avoid image-image interactions along the sheets normal direction. After obtaining the minimized structures, we applied loading conditions to evaluate the mechanical properties. For this purpose, the periodic simulation box size along the loading direction was increased in a multistep procedure, every step with a small engineering strain of 0.001. For the uniaxial loading conditions, upon the stretching along the loading direction the stress along the transverse direction should be negligible. To satisfy this condition, after applying the loading strain, the simulation box size along the transverse direction of the loading was changed accordingly in a way that the transverse stress remained negligible in comparison with the stress along the loading direction. For the biaxial loading condition, the equal loading strain was applied simultaneously along the both planar directions. After applying the changes in the simulation box size, the atomic positions were rescaled to avoid any sudden void formation or bond stretching as well. We then used the conjugate gradient method for the geometry optimizations, using the convergence criteria of $10^{-5}$ eV and 0.005 eV/Å for the energy and the forces, respectively. For the 2H and 1T phases we used a 15×15×1 Monkhorst-Pack [53] k-point mesh size whereas for the 1T' a 7×14×1 Monkhorst-Pack grid was used. To more precisely calculate the electronic density of states (DOS), we conducted single point calculations using the tetrahedron method with Blöchl corrections in which the Brillouin zone was sampled with a 25×25×1 Monkhorst k-point mesh size. Since the PBE functional usually underestimate the band-gap values, we also calculated the DOS using the HSE06 [54] hybrid functional with 15×15×1 k-point mesh size.

## 3. Results and discussions

We first study the atomic structures of energy minimized single-layer $MoTe_2$ pristine phases. Fig.1, illustrates the atomic lattice of 2H-, 1T- and 1T'-$MoTe_2$. Likely to graphene, $MoTe_2$ structures also consists of two major directions of armchair and zigzag, as shown in Fig. 1. The 2H structure in $MoTe_2$ shows the ABA atomic stacking sequence whereas the 1T and 1T' demonstrate the ABC atomic stacking sequence. For the 2H and 1T, the atomic lattice can be well defined by the hexagonal lattice constant (α) and the Mo-Te bond length. In this study, the lattice constants for the 2H and 1T phases were predicted to be 3.549 Å and 3.495 Å, respectively. The lattice constant of 2H-MoTe$_2$ was experimentally reported to be 3.551 Å, which is in less than 0.06% difference with our estimation. The Mo-Te bond length for the 2H- and 1T-MoTe$_2$ was also calculated to be 2.731 Å and 2.756 Å, respectively. For the 1T'-MoTe$_2$ structure, the unit-cell includes a rectangular lattice with dimensions of 3.449 Å and 6.374 Å, which match excellently with values of 3.452 Å and 6.368 Å reported in the recent experimental work by Wang et al. [29]. In 1T'-MoTe$_2$ two different Mo-Te bonds exist with lengths of 2.718 Å and 2.823 Å.

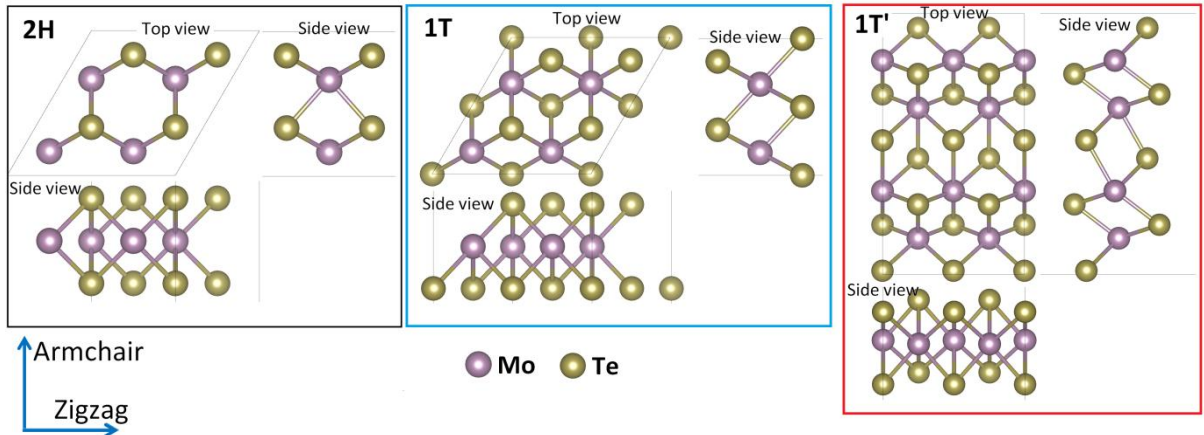

Fig. 1, Top and side views of pristine single-layer MoTe$_2$ with 2H, 1T and 1T' atomic lattices. We studied the mechanical properties along the armchair and zigzag directions as shown.

We next study the mechanical responses of these 2D structures by conducting the uniaxial tensile simulations. In Fig. 2, the DFT predictions for the uniaxial stress-strain responses of 2H, 1T and 1T'-MoTe$_2$, elongated along the armchair and zigzag directions are plotted. In all cases, the stress-strain responses present an initial linear relation which is followed by a nonlinear trend up to the ultimate tensile strength point, a point at which the structure yields its maximum load bearing strength. The slope of the first initial linear section of the stress-strain response is equal to the elastic modulus. We therefore fitted a line to the stress-strain values for the strain levels below 0.01 to report the elastic modulus. For these initial strain levels within the elastic limit, the strain along the traverse direction ($\varepsilon_t$) with respect to the loading strain ($\varepsilon_l$) is acceptably constant and can be used to obtain the Poisson's ratio using the $-\varepsilon_t/\varepsilon_l$. For the 2H and 1T single-layer $MoTe_2$ structures, the acquired results shown in Fig. 2 reveal that the initial linear responses match closely which accordingly suggest isotropic elastic properties. For the 1T'-$MoTe_2$ however along the armchair the stress values take higher values which confirm anisotropic elastic response. For the 2H and 1T single-layer $MoTe_2$, we predicted the elastic modulus of 76 N/m and 82 N/m, respectively. The Poisson's ratio was also estimated to be 0.25 and -0.03 for 2H and 1T lattices, respectively. Interestingly the 1T-$MoTe_2$ presents a negative Poisson's ratio and thus can be categorized as an auxetic material. The elastic modulus of 1T'-$MoTe_2$ along the armchair and zigzag directions were calculated to be 66 N/m and 43 N/m, respectively.

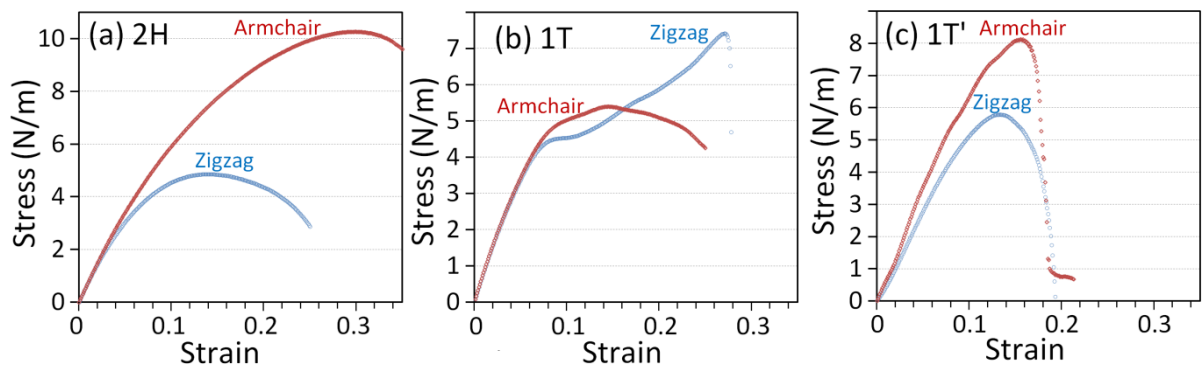

Fig. 2, Uniaxial stress-strain responses of polymorphic single-layer free-standing pristine $MoTe_2$ lattices stretched along the armchair and zigzag directions.

On the other hand, the non-linear part of the stress-strain curves of pristine $MoTe_2$ structures are found to be distinctly different depending on the loading direction. For the 2H- and 1T'-$MoTe_2$ the structures show higher tensile strength along the armchair direction compared with the zigzag one. Such an observation is in agreement with earlier studies for the mechanical properties of 2H transition metal dichalcogenides [19] and group IV graphene-like 2D materials. Nevertheless, this finding is reversed for 1T-MoTe$_2$ in which along the zigzag direction the structure yields a higher tensile strength and stretchability.

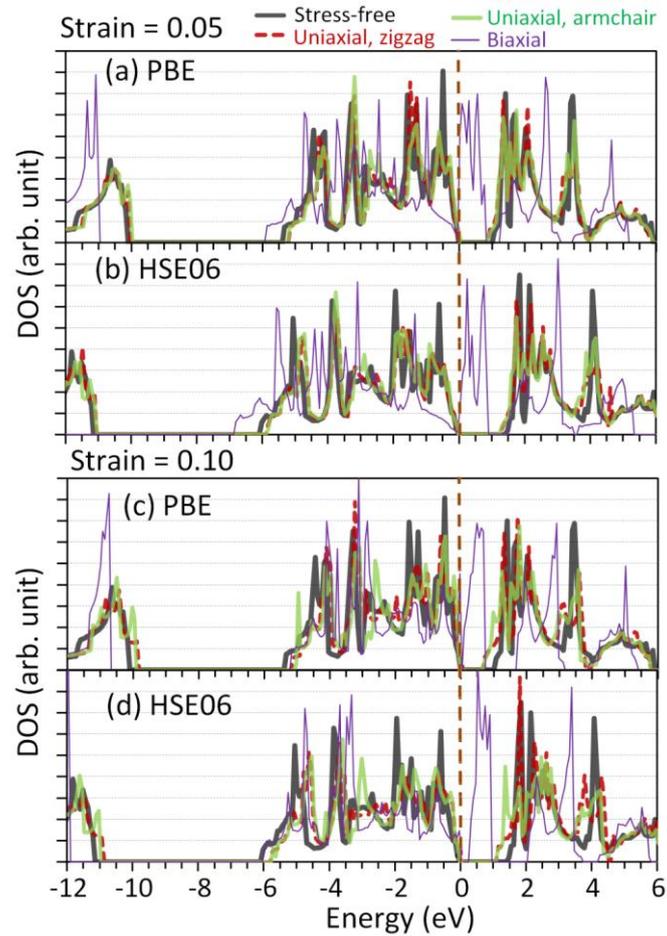

Fig. 3, Total DOS of unstrained and strained defect-free 2H-MoTe$_2$, for different biaxial or uniaxial strain levels of 0.05 and 0.10 as predicted by the PBE and HSE06 methods. The Fermi level is aligned to zero energy, and it is presented by the dashed brown line.

To better understand the underlying mechanism causing anisotropic tensile responses of MoTe$_2$ pristine phases, we analyzed the deformation process. For the deformation of 2H phase during the uniaxial loading, two different bonds can be distinguished, the bond along the armchair and the bond oriented along the zigzag direction. During the uniaxial tensile loading of single-layer 2H-MoTe$_2$, the bond oriented along the loading direction increases and the other bond oriented along the transverse direction of loading contracts. For the stretching of 2H-MoTe$_2$ along the armchair direction, one bond is exactly along the loading direction and directly involves in the load bearing and such that by increasing the strain level this bond increases substantially. In this case, the bond oriented along the transverse direction contract slightly. On the other side, for the uniaxial stretching along the zigzag, one bond is almost oriented along the loading direction and the other bond is exactly along the transverse direction. In this case, since the bond involving in the load transfer is not exactly inline of the loading by increasing the strain level the bond stretching is moderate. In this case the contraction of the bond along the transverse direction is more considerable, which helps the material to flow easier along the loading direction. As it is clear, the distinctly higher tensile strength of 2H-MoTe$_2$ along the armchair direction can be attributed to the fact that during the stretching along the armchair half of the bonds are exactly aligned to the loading direction and such that the deformation is achieved mainly by the bond elongation.

Table 1, Summary of mechanical properties of single-layer and pristine MoTe$_2$ for 2H, 1T and 1T' atomic configurations. E, P, STS and UTS stand for the elastic modulus, Poisson's ratio, strain at the ultimate tensile strength point and the ultimate tensile strength, respectively. The stress unit is in N/m, and subscripts zig. and arm. represent the properties along the zigzag and armchair directions, respectively.

| Structure | E$_{arm.}$ | E$_{zig.}$ | P$_{arm.}$ | P$_{zig.}$ | UTS$_{arm.}$ | UTS$_{zig.}$ | STS$_{arm.}$ | STS$_{zig.}$ |
|---|---|---|---|---|---|---|---|---|
| 2H | 76 | 75 | 0.25 | 0.25 | 10.3 | 4.9 | 0.3 | 0.14 |
| 1T | 82 | 82 | -0.03 | -0.03 | 5.4 | 7.4 | 0.15 | 0.27 |
| 1T' | 66 | 43 | 0.49 | 0.42 | 8.1 | 5.8 | 0.16 | 0.14 |

For the stretching of 1T-MoTe$_2$ along the armchair direction, from every three bonds, two bonds are oriented almost along the transverse direction of loading. In this case, only one Mo–Te bond, which is exactly along the loading direction is incorporated in the load bearing. Therefore, along the armchair direction the stretchability of this bond plays a crucial role in the mechanical response. In contrast for the loading along the zigzag direction, two bonds are almost oriented along the loading direction and the other bond is exactly along the transverse direction. This way, for the 1T-MoTe$_2$ nanomembranes stretched along the zigzag more bonds are involved in stretching and accordingly in the load transfer which can explain the higher tensile strength and stretchability as well. Worthy to note that 1T'-MoTe$_2$ shows the least anisotropy in the mechanical response in comparison with 2H and 1T phases. The predicted mechanical properties of pristine MoTe$_2$ phases are summarized in Table 1.

Next, we analyzed electronic properties of single-layer MoTe$_2$ pristine phases under different loading conditions. In Fig. 3, the calculated total DOS predicted by the PBE and HSE06 methods for 2H-MoTe$_2$ under different loading conditions are compared. For the unstrained 2H-MoTe$_2$ the band-gap was predicted to be ~0.7 eV and ~1.4 eV by PBE and HSE06 functional, respectively. As expected the PBE underestimates the electronic band-gap. According to our results depicted in Fig. 3, both PBE and HSE06 predict very similar trends, although the estimated band-gap values are different. For 2H-MoTe$_2$ the both PBE and HSE06 methods suggest very close band-gaps for the uniaxial loading along the armchair and zigzag directions and in the both cases by increasing the loading strain the band-gap decreases. For the biaxial loading at strain the level of 0.05 (Fig. 3a and Fig. 3b), both PBE and HSE06 approaches reveal that at the zero-state energy (Fermi level) the DOS is not zero which confirms that the metallic electronic character. By increasing the strain level to 0.10 the PBE approach yet suggests metallic response whereas the HSE06 shows a slight band-gap opening. As it is clear, the electronic properties of 2H-MoTe$_2$ yield a complicated behaviour for the biaxial loading while the uniaxial loading shows very promising performance for tuning the electronic band-gap.

Since the HSE06 is a more valid approach for studying the electronic properties such as the band-gap, we concentrated on this method results for studying the electronic properties of strained 1T and 1T'-MoTe$_2$. We remind that in agreement with previous study [26] we found that both unstrained 1T and 1T'-MoTe$_2$ exhibit metallic electronic response. In Fig. 3, the calculated total electronic DOS for 1T- and 1T'-MoTe$_2$ under different loading conditions are illustrated. For the 1T-MoTe$_2$ the obtained results suggest that the metallic electronic character was kept intact upon the different loading conditions and only for the biaxial loading with high strain level of 0.10 a slight band-gap opening of ~0.2 eV is observable. These results clearly suggest limited chance for the band-gap opening in 1T-MoTe$_2$ through applying the mechanical loadings. On the other side for the 1T'-MoTe$_2$, for the strain level of 0.05 both the uniaxial loading along the zigzag and biaxial loading could lead to opening of ~0.2 eV band-gap whereas for the uniaxial loading along the armchair the metallic character is retained. In this case, by increasing the strain level to 0.10 all strained systems show semiconducting electronic character with very narrow band-gaps.

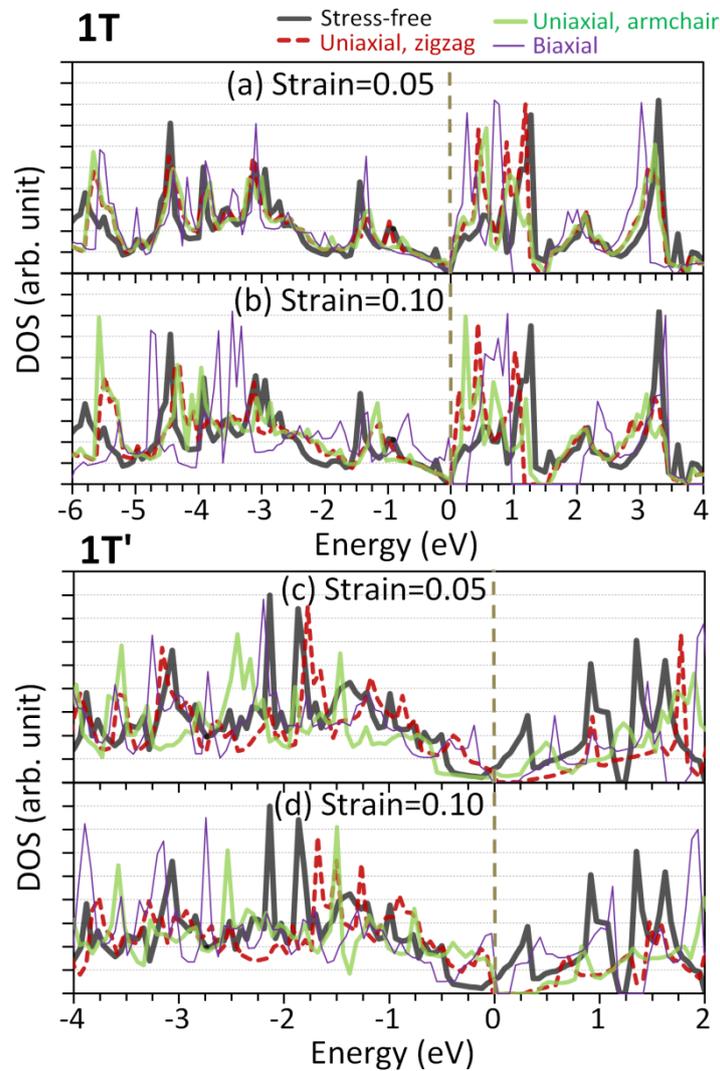

Fig. 4, Total electronic DOS of unstrained and strained 1T and 1T'-MoTe$_2$, predicted by the HSE06 functional. The Fermi level is aligned to zero energy, and it is presented by the dashed brown line.

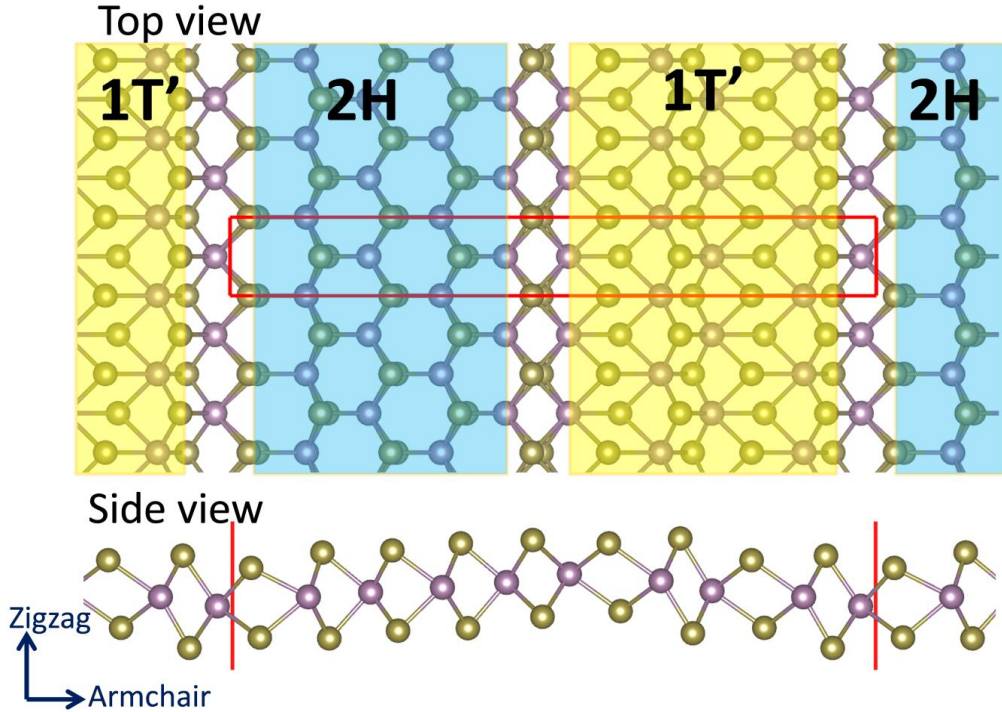

Fig. 5, Top and side view of the studied 1T'/2H-MoTe$_2$ heterostructure which includes 9 Mo and 18 Te atoms. The Red box indicates the unit cell boundaries.

We finally analyze the strength of 1T'/2H-MoTe$_2$ heterostructure. In Fig. 5, top and side views of the 1T'/2H-MoTe$_2$ heterostructure studied in this investigation are depicted, which include 9 Mo and 18 Te atoms. To build the interfaces between 2H- and 1T'-MoTe$_2$, observing the periodicity of the structure and also taking into consideration of computational limits of the DFT simulations, two different boundaries were formed. In this regard, the first interface which exists on the middle of the constructed structure is exactly on the basis of experimental observation by Sung *et al.* [28]. This interface includes a lozenge geometry formed by 4 Te atoms bridging the 2H and 1T' phases. The next interface that we found that may exists in experimental samples is shown on the boundary of the constructed system in Fig. 5, which includes also a lozenge geometry but made from a single Mo atom stitching the two different MoTe$_2$ phases. As it is clear the first 1T'/2H interface shows a denser configuration. Moreover, as observable in the side view of constructed structure shown in Fig. 5, a slight bucking occurs as a result of existing interfaces in the structure. This can be attributed to the fact that we considered limited number of atoms in this system because of the computational costs of DFT simulations.

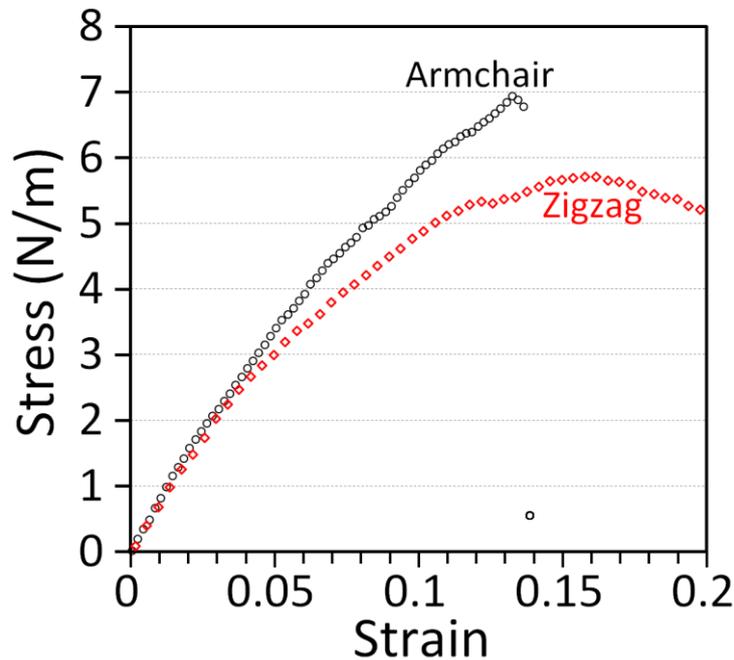

Fig. 6, Uniaxial stress-strain response of the single-layer 1T'/2H-MoTe$_2$ heterostructure for the loading along the normal (armchair) and parallel (zigzag) direction of 1T'/2H interface.

In this work, we investigate the mechanical properties of 1T'/2H-MoTe$_2$ heterostructure by conducting the uniaxial tensile simulations along the normal (armchair) and parallel (zigzag) directions of 1T'/2H interface (as depicted in Fig. 5). In Fig. 6, the acquired DFT results for the uniaxial stress-strain responses of 1T'/2H-MoTe$_2$ heterostructure are illustrated. Likely to the pristine MoTe$_2$ phases, the stress-strain response of 1T'/2H heterostructure starts with an initial linear relation followed by a non-linear curve up to the ultimate tensile strength point. In comparison with the pristine 1T'- and 2H-MoTe$_2$ elongated along the armchair direction, the 1T'/2H heterostructure along the armchair direction exhibits lower stretchability (with a strain at rupture of ~0.14). Moreover, the tensile strength of heterostructure when stretched along the normal direction of 1T'/2H interface is found to be below than that of the both native phases along the armchair direction. These preliminary observations can be counted as signs that in the constructed heterostructure the rupture occurred along the 1T'/2H interface, when loaded along the normal direction of the formed interface. In Table 2 we compare the tensile strengths of 2H and 1T' pristine phases and 1T'/2H heterostructure for the uniaxial tensile loading along the armchair and zigzag directions. Interestingly, for the uniaxial loading along the zigzag direction, the tensile strength of 1T'/2H heterostructure is higher than that of the native 2H phase along the zigzag direction. These results accordingly confirm the remarkable mechanical strength of 1T'/2H-MoTe$_2$ heterostructures.

Table 2, Comparison of ultimate tensile strengths of 2H, 1T' and 1T'/2H heterostructure for the uniaxial tensile loading along the armchair and zigzag directions.

| Structure | Tensile strength (N/m) | |
| --- | --- | --- |
|  | Armchair | Zigzag |
| 1T'/2H heterostructure | 7.0 | 5.7 |
| 2H | 10.3 | 4.9 |
| 1T' | 8.1 | 5.8 |

In Fig. 7, the deformation process of 1T'/2H-MoTe$_2$ heterostructure is illustrated. As it is clear, the rupture in the Mo-Te bond happens for the 1T'/2H interface existing at the boundary of system. Interestingly, the 1T'/2H interface that was experimentally realized by Sung *et al.* [28], was found to be intact which confirms its higher tensile strength. Since the ruptured interface is not the one reported in the experimentally fabricated 1T'/2H-MoTe$_2$ heterostructure, our predicted tensile strength of 7 N/m should be considered as a lower bound and such that the real experimental samples may endure at higher stress values. This observation further confirms the outstanding mechanical properties of 1T'/2H-MoTe$_2$ heterostructures. Our deformation analysis also shows that by applying the strain levels the buckling of the constructed model only slightly changes which reveal that the deformation is mainly achieved by the bond stretching. Worthy to remind that according to the experimental observations by Sung *et al.* [28], the 1T'/2H interfaces were found to form mainly along the zigzag edges, which can be explained by the minimal mismatch in the atomic lattices of the two native phases in these cases. Nevertheless, further theoretical and experimental investigations on the formations of different types of interfaces along the MoTe$_2$ heterostructures can be an interesting topic for the future studies.

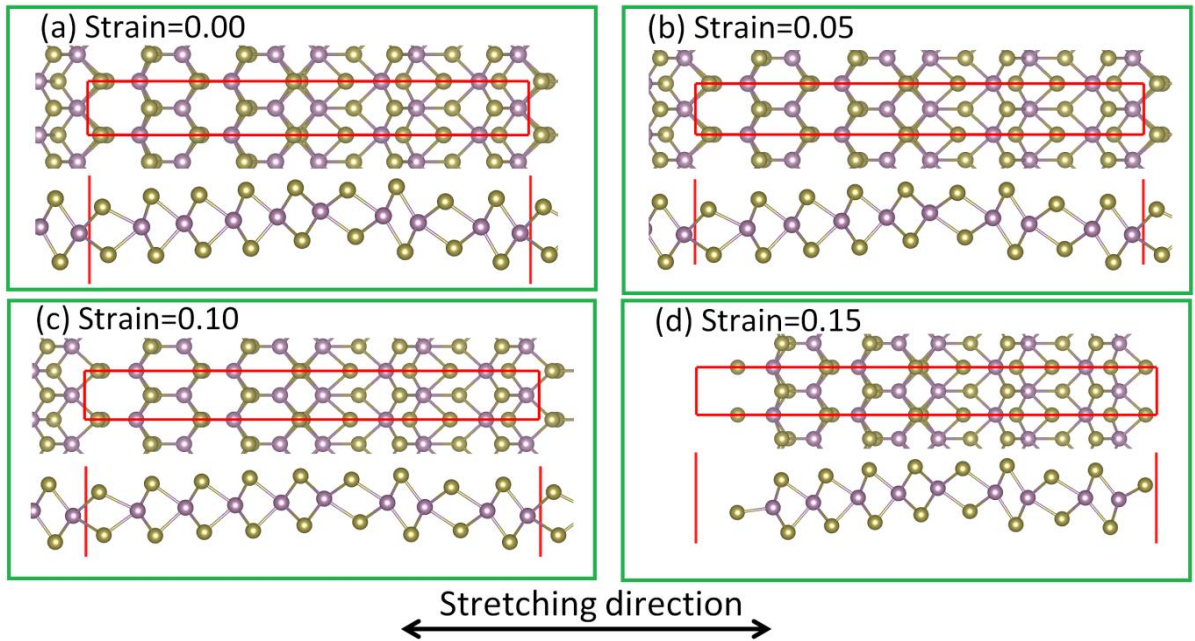

Fig. 7, Top and side views of uniaxial tensile deformation processes of a single-layer 1T'/2H-MoTe$_2$ heterostructure at different strain levels. The structure is stretched along the normal direction of the 1T'/2H interface (armchair direction).

## 4. Conclusion

The emerging two-dimensional transition-metal dichalcogenides offer appealing physical and chemical properties. In this regard, MoTe$_2$ nanosheets have recently attracted remarkable attention because of their very promising structural, electronic, photonic and optical properties. MoTe$_2$ interestingly shows polymorphism nature, which may exist in three different structural phases of 2H, 1T and 1T', offering diverse physical and chemical properties. Most recently, 1T'/2H-MoTe$_2$ a novel coplanar heterostructure was experimentally realized, which shows promising performances as a field-effect transistor. In this study, we accordingly conducted first-principles density functional theory calculations to explore the mechanical properties of single-layer MoTe$_2$ structures. We first studied the mechanical responses of pristine 2H, 1T and 1T'-MoTe$_2$ structures by conducting the uniaxial tensile simulations. Our modelling results reveal that the mechanical responses of these two-dimensional systems are not isotropic and the atomic configurations in the MoTe$_2$ lattices can considerably influence the tensile properties. It was found that the 1T-MoTe$_2$ presents a negative Poisson's ratio and thus can be categorized as an auxetic material. Based on the PBE and HSE06 calculations, it was found that for 2H-MoTe$_2$ the band-gap decreases through applying uniaxial tensile loading, which notably confirms the tunability of electronic response of this 2D structure. The acquired results nevertheless suggest limited support for the band-gap opening in 1T and 1T'-MoTe$_2$ through applying the mechanical loadings. Our analysis of the mechanical properties of 1T'/2H-MoTe$_2$ heterostructure confirms that this novel two-dimensional structure can yield a remarkably high strength. The insights provided by this study can be useful for the practical application of various forms of MoTe$_2$ nanomembranes including homo and heterostructures for new generation nanodevices.

**Acknowledgment**

B. M. and T. R. greatly acknowledge the financial support by European Research Council for COMBAT project (Grant number 615132).

**References**

(1) Novoselov, K. S.; Geim, A. K.; Morozov, S. V; Jiang, D.; Zhang, Y.; Dubonos, S. V; Grigorieva, I. V; Firsov, A. A. Electric Field Effect in Atomically Thin Carbon Films. *Science* **2004**, *306*, 666–669.
(2) Geim, A. K.; Novoselov, K. S. The Rise of Graphene. *Nat. Mater.* **2007**, *6*, 183–191.
(3) Guinea, F.; Katsnelson, M. I.; Geim, a. K. Energy Gaps, Topological Insulator State and Zero-Field Quantum Hall Effect in Graphene by Strain Engineering. *Nat. Phys.* **2009**, *6*, 30–33.
(4) Lherbier, A.; Blase, X.; Niquet, Y. M.; Triozon, F.; Roche, S. Charge Transport in Chemically Doped 2D Graphene. *Phys. Rev. Lett.* **2008**, *101*.
(5) Martins, T. B.; Miwa, R. H.; Da Silva, A. J. R.; Fazzio, A. Electronic and Transport Properties of Boron-Doped Graphene Nanoribbons. *Phys. Rev. Lett.* **2007**, *98*.
(6) Geim, a K.; Grigorieva, I. V. Van Der Waals Heterostructures. *Nature* **2013**, *499*, 419–425.
(7) Wang, Q. H.; Kalantar-Zadeh, K.; Kis, A.; Coleman, J. N.; Strano, M. S. Electronics and Optoelectronics of Two-Dimensional Transition Metal Dichalcogenides. *Nat. Nanotechnol.* **2012**, *7*, 699–712.

(8)  Radisavljevic, B.; Radenovic, A.; Brivio, J.; Giacometti, V.; Kis, A. Single-Layer MoS$_2$ Transistors. *Nat. Nanotechnol.* **2011**, *6*, 147–150.

(9)  Presolski, S.; Pumera, M. Covalent Functionalization of MoS2. *Mater. Today* **2016**, *19*, 140–145.

(10) Coleman, J. N.; Lotya, M.; O'Neill, A.; Bergin, S. D.; King, P. J.; Khan, U.; Young, K.; Gaucher, A.; De, S.; Smith, R. J.; *et al.* Two-Dimensional Nanosheets Produced by Liquid Exfoliation of Layered Materials. *Science (80-. ).* **2011**, *331*, 568–571.

(11) Xenogiannopoulou, E.; Tsipas, P.; Aretouli, K. E.; Tsoutsou, D.; Giamini, S. A.; Bazioti, C.; Dimitrakopulos, G. P.; Komninou, P.; Brems, S.; Huyghebaert, C.; *et al.* High-Quality, Large-Area MoSe2 and MoSe2/Bi2Se3 Heterostructures on AlN(0001)/Si(111) Substrates by Molecular Beam Epitaxy. *Nanoscale* **2015**, *7*, 7896–7905.

(12) Lee, Y. H.; Zhang, X. Q.; Zhang, W.; Chang, M. T.; Lin, C. T.; Chang, K. D.; Yu, Y. C.; Wang, J. T.; Chang, C. S.; Li, L. J.; *et al.* Synthesis of Large-Area MoS2 Atomic Layers with Chemical Vapor Deposition. *Adv Mater* **2012**, *24*, 2320–2325.

(13) Radisavljevic, B.; Radenovic, A.; Brivio, J.; Giacometti, V.; Kis, A. Single-Layer MoS2 Transistors. *Nat. Nanotechnol.* **2011**, *6*, 147–150.

(14) Splendiani, A.; Sun, L.; Zhang, Y.; Li, T.; Kim, J.; Chim, C. Y.; Galli, G.; Wang, F. Emerging Photoluminescence in Monolayer MoS2. *Nano Lett.* **2010**, *10*, 1271–1275.

(15) Lopez-Sanchez, O.; Lembke, D.; Kayci, M.; Radenovic, A.; Kis, A. Ultrasensitive Photodetectors Based on Monolayer MoS2. *Nat. Nanotechnol.* **2013**, *8*, 497–501.

(16) Ostadhossein, A.; Yoon, K.; van Duin, A. C. T.; Seo, J. W.; Seveno, D. Do Nickel and Iron Catalyst Nanoparticles Affect the Mechanical Strength of Carbon Nanotubes? *Extrem. Mech. Lett.*

(17) Conley, H. J.; Wang, B.; Ziegler, J. I.; Haglund, R. F.; Pantelides, S. T.; Bolotin, K. I. Bandgap Engineering of Strained Monolayer and Bilayer MoS2. *Nano Lett.* **2013**, *13*, 3626–3630.

(18) Castellanos-Gomez, A.; Roldán, R.; Cappelluti, E.; Buscema, M.; Guinea, F.; Van Der Zant, H. S. J.; Steele, G. A. Local Strain Engineering in Atomically Thin MoS2. *Nano Lett.* **2013**, *13*, 5361–5366.

(19) Li, J.; Medhekar, N. V.; Shenoy, V. B. Bonding Charge Density and Ultimate Strength of Monolayer Transition Metal Dichalcogenides. *J. Phys. Chem. C* **2013**, *117*, 15842–15848.

(20) Kou, L.; Frauenheim, T.; Chen, C. Nanoscale Multilayer Transition-Metal Dichalcogenide Heterostructures: Band Gap Modulation by Interfacial Strain and Spontaneous Polarization. *J. Phys. Chem. Lett.* **2013**, *4*, 1730–1736.

(21) Guzman, D. M.; Strachan, A. Role of Strain on Electronic and Mechanical Response of Semiconducting Transition-Metal Dichalcogenide Monolayers: An Ab-Initio Study. *J. Appl. Phys.* **2014**, *115*.

(22) Chhowalla, M.; Shin, H. S.; Eda, G.; Li, L.-J.; Loh, K. P.; Zhang, H. The Chemistry of Two-Dimensional Layered Transition Metal Dichalcogenide Nanosheets. *Nat. Chem.* **2013**, *5*, 263–275.

(23) He, K.; Poole, C.; Mak, K. F.; Shan, J. Experimental Demonstration of Continuous Electronic Structure Tuning via Strain in Atomically Thin MoS2. *Nano Lett.* **2013**, *13*, 2931–2936.

(24) Chhowalla, M.; Voiry, D.; Yang, J.; Shin, H. S.; Loh, K. P. Phase-Engineered Transition-Metal Dichalcogenides for Energy and Electronics. *MRS Bull.* **2015**, *40*, 585–591.

(25) Calandra, M. Chemically Exfoliated Single-Layer MoS 2: Stability, Lattice Dynamics, and Catalytic Adsorption from First Principles. *Phys. Rev. B - Condens. Matter Mater.*

*Phys.* **2013**, *88*.

(26) Qian, X.; Liu, J.; Fu, L.; Li, J. Quantum Spin Hall Effect and Topological Field Effect Transistor in Two - Dimensional T Ransition Metal Dichalcogenides. *Science (80-. ).* **2014**, *346*, 1344–1347.

(27) Jariwala, D.; Sangwan, V. K.; Lauhon, L. J.; Marks, T. J.; Hersam, M. C. Emerging Device Applications for Semiconducting Two-Dimensional Transition Metal Dichalcogenides. *ACS Nano*, 2014, *8*, 1102–1120.

(28) Sung, J. H.; Heo, H.; Si, S.; Kim, Y. H.; Noh, H. R.; Song, K.; Kim, J.; Lee, C.-S.; Seo, S.-Y.; Kim, D.-H.; *et al.* Coplanar Semiconductor–metal Circuitry Defined on Few-Layer MoTe2 via Polymorphic Heteroepitaxy. *Nat. Nanotechnol.* **2017**, *12*, 1064.

(29) Wang, Y.; Xiao, J.; Zhu, H.; Li, Y.; Alsaid, Y.; Fong, K. Y.; Zhou, Y.; Wang, S.; Shi, W.; Wang, Y.; *et al.* Structural Phase Transition in Monolayer MoTe2 Driven by Electrostatic Doping. *Nature* **2017**, *550*, 487.

(30) Bie, Y.-Q.; Grosso, G.; Heuck, M.; Furchi, M. M.; Cao, Y.; Zheng, J.; Bunandar, D.; Navarro-Moratalla, E.; Zhou, L.; Efetov, D. K.; *et al.* A MoTe2-Based Light-Emitting Diode and Photodetector for Silicon Photonic Integrated Circuits. *Nat. Nanotechnol.* **2017**, *12*, 1124.

(31) Lin, Y.-C.; Dumcenco, D. O.; Huang, Y.-S.; Suenaga, K. Atomic Mechanism of the Semiconducting-to-Metallic Phase Transition in Single-Layered MoS2. *Nat. Nanotechnol.* **2014**, *9*, 391–396.

(32) Empante, T. A.; Zhou, Y.; Klee, V.; Nguyen, A. E.; Lu, I. H.; Valentin, M. D.; Naghibi Alvillar, S. A.; Preciado, E.; Berges, A. J.; Merida, C. S.; *et al.* Chemical Vapor Deposition Growth of Few-Layer MoTe2 in the 2H, 1T′, and 1T Phases: Tunable Properties of MoTe2 Films. *ACS Nano* **2017**, *11*, 900–905.

(33) Akinwande, D.; Brennan, C. J.; Bunch, J. S.; Egberts, P.; Felts, J. R.; Gao, H.; Huang, R.; Kim, J.; Li, T.; Li, Y.; *et al.* A Review on Mechanics and Mechanical Properties of 2D Materials - Graphene and Beyond. *Extrem. Mech. Lett.* **2016**, *13*, 42–77.

(34) Ahmadpoor, F.; Sharma, P. A Perspective on the Statistical Mechanics of 2D Materials. *Extreme Mechanics Letters*, 2017, *14*, 38–43.

(35) Xiong, S.; Cao, G. Continuum Thin-Shell Model of the Anisotropic Two-Dimensional Materials: Single-Layer Black Phosphorus. *Extrem. Mech. Lett.* **2017**, *15*, 1–9.

(36) Hasanian, M.; Lissenden, C. J. Second Order Harmonic Guided Wave Mutual Interactions in Plate: Vector Analysis, Numerical Simulation, and Experimental Results. *J. Appl. Phys.* **2017**, *122*, 84901.

(37) Izadifar, M.; Abadi, R.; Jam, A. N.; Rabczuk, T. Investigation into the Effect of Doping of Boron and Nitrogen Atoms in the Mechanical Properties of Single-Layer Polycrystalline Graphene. *Comput. Mater. Sci.* **2017**, *138*, 435–447.

(38) Abadi, R.; Uma, R. P.; Izadifar, M.; Rabczuk, T. Investigation of Crack Propagation and Existing Notch on the Mechanical Response of Polycrystalline Hexagonal Boron-Nitride Nanosheets. *Comput. Mater. Sci.* **2017**, *131*, 86–99.

(39) Abadi, R.; Uma, R. P.; Izadifar, M.; Rabczuk, T. The Effect of Temperature and Topological Defects on Fracture Strength of Grain Boundaries in Single-Layer Polycrystalline Boron-Nitride Nanosheet. *Comput. Mater. Sci.* **2016**, *123*, 277–286.

(40) Sadeghzadeh, S. Borophene Sheets with in-Plane Chain-like Boundaries; a Reactive Molecular Dynamics Study. *Comput. Mater. Sci.* **2018**, *143*, 1–14.

(41) Alian, A. R.; Meguid, S. A.; Kundalwal, S. I. Unraveling the Influence of Grain Boundaries on the Mechanical Properties of Polycrystalline Carbon Nanotubes. *Carbon N. Y.* **2017**, *125*, 180–188.

(42) Alian, A. R.; Meguid, S. A. Molecular Dynamics Simulations of the Effect of Waviness and Agglomeration of CNTs on Interface Strength of Thermoset

(43) Jia, Z.; Li, T. Failure Mechanics of a Wrinkling Thin Film Anode on a Substrate under Cyclic Charging and Discharging. *Extrem. Mech. Lett.* **2016**, *8*, 273–282.

(44) Sha, Z.-D.; Pei, Q.-X.; Zhou, K.; Dong, Z.; Zhang, Y.-W. Temperature and Strain-Rate Dependent Mechanical Properties of Single-Layer Borophene. *Extrem. Mech. Lett.* **2018**, *19*, 39–45.

(45) Kan, M.; Nam, H. G.; Lee, Y. H.; Sun, Q. Phase Stability and Raman Vibration of the Molybdenum Ditelluride ($MoTe_2$) Monolayer. *Phys. Chem. Chem. Phys.* **2015**, *17*, 14866–14871.

(46) Bera, A.; Singh, A.; Muthu, D. V. S.; Waghmare, U. V.; Sood, A. K. Pressure-Dependent Semiconductor to Semimetal and Lifshitz Transitions in 2H-MoTe2: Raman and First-Principles Studies. *J. Phys. Condens. Matter* **2017**, *29*.

(47) Kumar, A.; Ahluwalia, P. K. Mechanical Strain Dependent Electronic and Dielectric Properties of Two-Dimensional Honeycomb Structures of MoX2 (X=S, Se, Te). *Phys. B Condens. Matter* **2013**, *419*, 66–75.

(48) Kresse, G. From Ultrasoft Pseudopotentials to the Projector Augmented-Wave Method. *Phys. Rev. B* **1999**, *59*, 1758–1775.

(49) Kresse, G.; Furthm??ller, J. Efficiency of Ab-Initio Total Energy Calculations for Metals and Semiconductors Using a Plane-Wave Basis Set. *Comput. Mater. Sci.* **1996**, *6*, 15–50.

(50) Kresse, G.; Furthmüller, J. Efficient Iterative Schemes for Ab Initio Total-Energy Calculations Using a Plane-Wave Basis Set. *Phys. Rev. B* **1996**, *54*, 11169–11186.

(51) Perdew, J.; Burke, K.; Ernzerhof, M. Generalized Gradient Approximation Made Simple. *Phys. Rev. Lett.* **1996**, *77*, 3865–3868.

(52) Liu, F.; Ming, P.; Li, J. Ab Initio Calculation of Ideal Strength and Phonon Instability of Graphene under Tension. *Phys. Rev. B - Condens. Matter Mater. Phys.* **2007**, *76*.

(53) Monkhorst, H.; Pack, J. Special Points for Brillouin Zone Integrations. *Phys. Rev. B* **1976**, *13*, 5188–5192.

(54) Scuseria, A. V. K. and O. A. V. and A. F. I. and G. E. Influence of the Exchange Screening Parameter on the Performance of Screened Hybrid Functionals. *J. Chem. Phys.* **2006**, *125*, 224106.